\documentclass[aps,prl,10pt,twocolumn,groupedaddress]{revtex4}

\usepackage{graphicx}
\usepackage{hyperref}
\usepackage{amsmath}
\usepackage{amsfonts}
\usepackage{amssymb}
\usepackage{subfigure}
\usepackage{times}

\begin{document}


\title{Real-Time Imaging of K atoms on Graphite: Interactions and Diffusion}


\author{J. Renard}
\email{renardj@phas.ubc.ca}
\affiliation{Department of Physics and Astronomy, University of British Columbia, Vancouver, British Columbia, V6T1Z4, Canada}
\author{M. B. Lundeberg}
\affiliation{Department of Physics and Astronomy, University of British Columbia, Vancouver, British Columbia, V6T1Z4, Canada}
\author{J. A. Folk}
\affiliation{Department of Physics and Astronomy, University of British Columbia, Vancouver, British Columbia, V6T1Z4, Canada}
\author{Y. Pennec}
\affiliation{Department of Physics and Astronomy, University of British Columbia, Vancouver, British Columbia, V6T1Z4, Canada}

\date{\today}

\begin{abstract}

Scanning tunneling microscopy (STM) at liquid helium temperature is used to image potassium adsorbed on graphite at low coverage ($\approx$0.02 monolayer). Single atoms appear as protrusions on STM topographs. A statistical analysis of the position of the atoms demonstrates repulsion between adsorbates, which is quantified by comparison with molecular dynamics simulations.  This gives access to the dipole moment of a single adsorbate, found to be 10.5$\pm$1 Debye. Time lapse imaging shows that long range order is broken by thermally activated diffusion, with a 32 meV barrier to hopping between graphite lattice sites. 

\end{abstract}


\maketitle

The addition of alkali atoms onto and into graphite has long been recognized as a powerful technique for engineering its electronic properties. Beyond the simple doping that results from a charge transfer, alkali intercalation has been shown to lead to superconductivity in several graphite compounds \cite{han65,wel05}. The emergence of graphene brought a renewed interest in the field: possibilities range from simply achieving very high doping levels\cite{che08}, to so-called atomic collapse for adatoms that transfer sufficient charge to the substrate \cite{shy07}, to inducing two-dimensional superconductivity in doped graphene \cite{uch07,mcc10}.

For all the recent excitement about alkali-on-graphite/ene systems, little is known about interactions between the adatoms and the surface, or between the adatoms themselves.  These are questions that must be resolved to better understand the effects of alkali doping. Coulomb interactions between adatoms are influenced by screening in the graphite, which itself is a topic of great interest as it relates to electron-electron interactions in this system. An understanding of adatom interactions is also important to the study of phase transitions in two dimensional systems, and of adsorbate self-organization phenomena on surfaces \cite{bar05,sch07}.

The structural phase diagram of alkali adatoms on graphite has previously been explored using low energy electron diffraction (LEED)\cite{li91}. For the case of potassium, a coverage higher than 0.1 monolayer (ML) leads to the condensation of a metallic 2x2 phase. In the dilute limit more commonly used for electronic doping of surfaces, K adatoms form a dispersed phase, with a large spacing that is believed to result from electrostatic repulsion between the partially-ionized adatoms after charge transfer to the graphite\cite{li91}.  Although spectroscopic and structural studies (LEED, Electron Energy Loss Spectroscopy, Photoemission) support this general picture \cite{li91,ost99,alg06}, local probes have the potential to offer much more detailed insight into adatom interactions \cite{yin09}. 

In this letter we take advantage of the real space imaging capability of the scanning tunneling microscope (STM) to explore the spatial distribution and dynamics of potassium adsorbed on graphite in the dilute limit.   Images of single stationary K atoms on highly oriented pyrolytic graphite (HOPG) are obtained by cooling the sample to 11K and operating at large tip-sample distances---the first time such a measurement has been performed despite the long history of interest in the K-on-graphite system.  The pair distribution function extracted from the spatial arrangement of K atoms deviates strongly from a random distribution, although no long-range order is observed. Quantitative comparison to molecular dynamics simulations suggests that K-on-graphite carry a dipole moment of 10.5$\pm$1 Debye due to the charge transfer. The expected hexagonal rotational order for such isotropic repulsive interaction is broken by thermally activated diffusion. Time lapse imaging and atom tracking techniques allow diffusion trajectories to be mapped out in real time.  From the diffusion rate, an energy barrier of 32 meV is estimated for K hopping between lattice sites.

\begin{figure}
\subfigure{\label{fig1a}}
\subfigure{\label{fig1b}}
\includegraphics{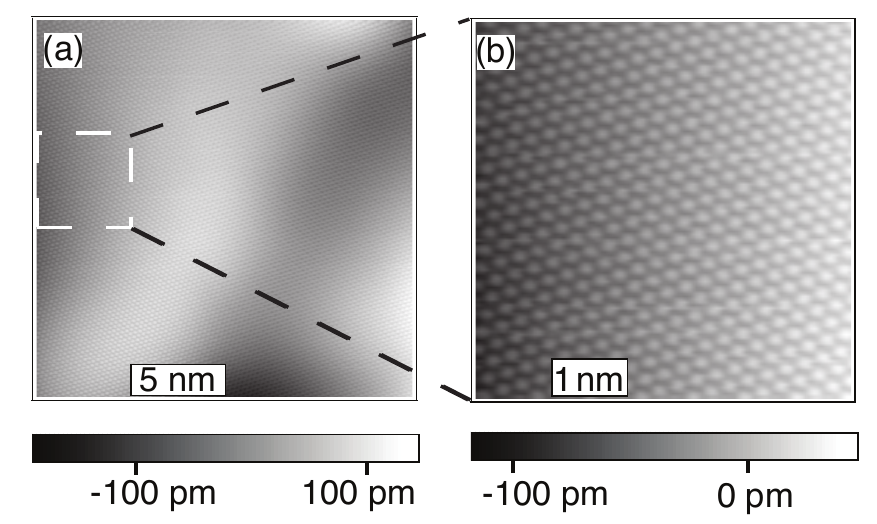}
\caption{\label{fig1}Topography of the HOPG graphite surface prior to potassium deposition. \subref{fig1a} Large scale image showing the clean surface. \subref{fig1b} Atomic resolution image of the area highlighted in \subref{fig1a}.The tunneling bias and current are 100 mV and 100 pA. White rectangles are 5 nm (a) and 1 nm (b) scale bars. }
\end{figure}

Measurements were performed under ultra high vacuum conditions ($\approx10^{-10}$mbar) at 11 K. A sample of HOPG was cleaved in-situ to avoid any possible surface contamination from exposition to ambient atmosphere.  A beetle type STM provided scanned images, using a W tip prepared in-situ through argon sputtering and high temperature annealing.  A typical surface obtained after cleaving can be seen in Fig.~\ref{fig1a}: large atomically flat terraces free of defects impurities or surface contaminant confirm the instrument stability and surface preparation protocol. Figure \ref{fig1b} zooms in a smaller region with atomic resolution, showing the triangular lattice characteristic of Bernal stacking where only one out of two atoms is observed \cite{won09}.

\begin{figure}
\subfigure{\label{fig2a}}
\subfigure{\label{fig2b}}
\subfigure{\label{fig2c}}
\includegraphics{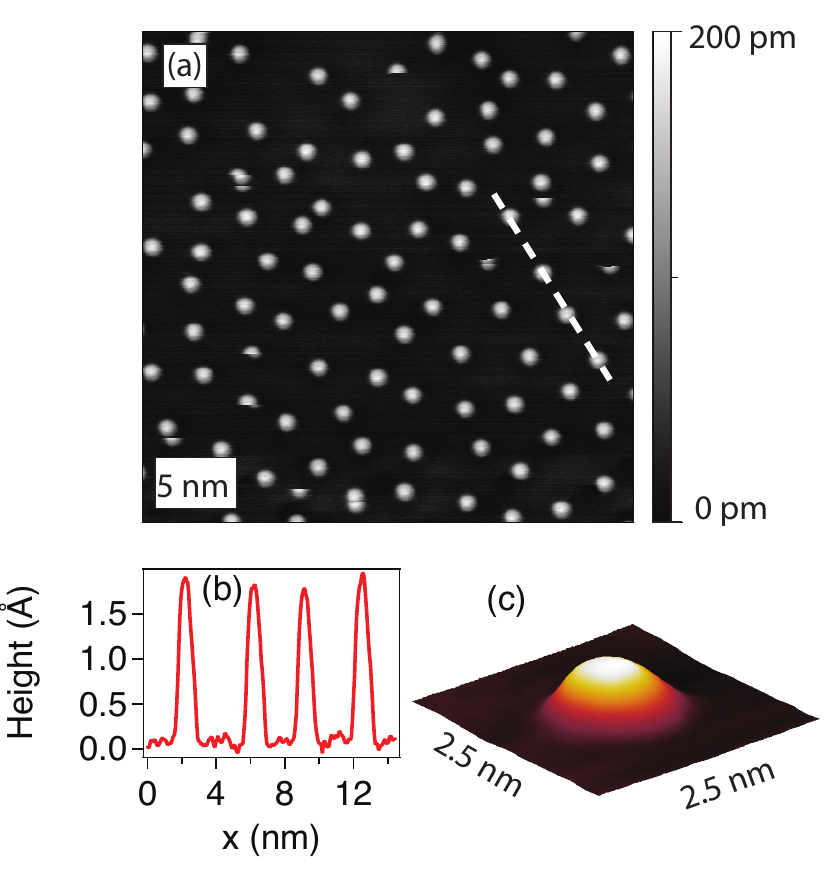}
\caption{\label{fig2} \subref{fig2a} Topographic image of the graphite surface after the depositon of potassium, showing isolated potassium atoms. The tunneling bias and current are -2V and 5 pA.   \subref{fig2b} The height profile  is taken along the dotted line drawn in \subref{fig2a}. \subref{fig2c} 3D landscape (full height 200pm) of a single potassium atom.}
\end{figure}

Potassium deposition was carried out with a 20 sec dose from an SAES getter heated with 6.5A of current and located 50 cm from the sample.  After K deposition, the tip-sample distance was greatly increased compared to the standard tunnelling conditions used for Fig.~\ref{fig1}, in order to image the potassium atoms without perturbing them with the tip (dragging or pushing them around) . This was accomplished by operating at a high bias voltage (typically -2 V) and very low current (typically 5 pA)---the equivalent of 400 G$\Omega$ impedance compared to 1 G$\Omega$ in Fig.~\ref{fig1}. Under those extreme tunnelling conditions, single potassium atoms are clearly observed as bright spots on the graphite sample  (Fig.~\ref{fig2}). The drawback of the large tip-sample distance is the loss of atomic resolution in the substrate image.  Calculations and LEED experiments indicate that K adatoms sit on the empty sites in the middle of a graphite honeycomb lattice cell \cite{fer04, anc93,ryt07,cha08}, but this could not be confirmed due to the lack of atomic resolution in the graphite while imaging K atoms.

A linescan through Fig.~\ref{fig2a} indicates an effective height for the K atoms of 2 \AA\ and a width of 7.8 \AA\ (full width at half maximum), reminiscent of single adatoms on metallic surfaces \cite{eig90,eig91,cro93}. By comparison, calculations and LEED measurements of the crystallographic height (center of the K adsorbate to center of the C surface atoms) suggest an actual height of 2.7 \AA \cite{fer04,ryt07,cha08}.  The discrepancy between these values may be explained by the fact than an STM measurement convolves the crystallographic height with tunnelling density of states integrated over bias. This explanation would imply that the density of states is smaller on potassium atoms than on the graphite surface.

\begin{figure}
\subfigure{\label{fig3a}}
\subfigure{\label{fig3b}}
\subfigure{\label{fig3c}}
\includegraphics{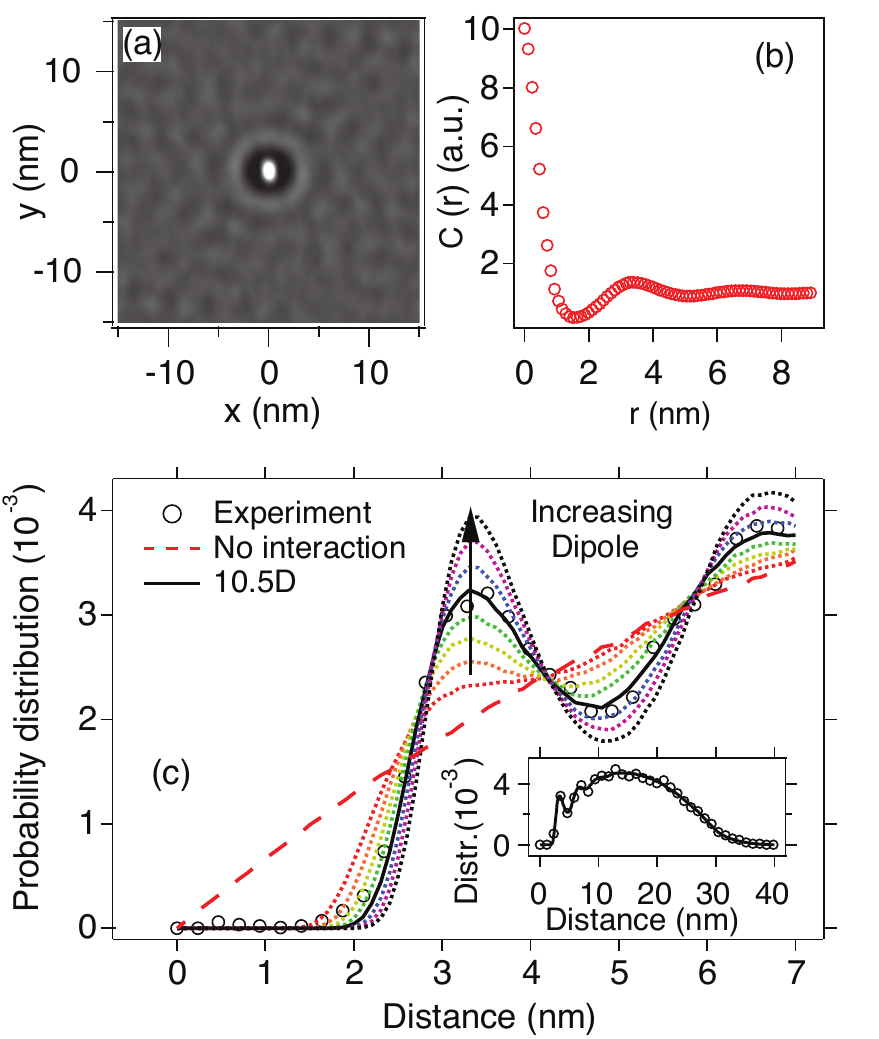}
\caption{\label{fig3}Study of the spatial organization of the diluted phase of the potassium atoms on graphite. \subref{fig3a} Autocorrelation image C(x,y) averaged on about 30 images similar to figure \ref{fig2a}. The dark area around the central peak evidences the repulsive interaction between the potassium atoms. \subref{fig3b} radial distribution function C(r), i.e average over the angle of C(x,y). \subref{fig3c} Pair distribution function, i.e., probability of having two atoms separated by a given distance. The empty circles are the experimental data; dotted lines are molecular dynamics simulations for evenly spaced dipole moment $\{6.2$D$..16.3$D$\}$, with a solid line for $10.5$D; dashed line is the expected distribution for a system without interaction, i.e. setting the dipole moment to 0 in the simulations. The inset shows the data and the best fit on a larger scale to show the effect of the finite size of the image.}
\end{figure}

Topographic images like those in Fig.~\ref{fig2a} provide information about K-K interactions.  It has been shown that graphite's two-dimensional and semi-metallic character strongly affect the spatial arrangement of potassium atoms on the surface, especially when compared to a metallic substrate \cite{ish92}. At first glance the distribution of atoms in Fig.~\ref{fig2a} may appear random, but statistical analysis demonstrates repulsion between adatoms. This is shown in Fig.~\ref{fig3a}  using a two dimensional autocorrelation of STM topographs, where the dark area around the center indicates repulsion between potassium adatoms.  For larger separations a rapidly decaying oscillatory behavior is observed, similar to what one might expect for a liquid with no long range orientational order.  This structure is typical for particles subject to a long range repulsive interaction in two dimensions \cite{lin07}, and consistent with the diffraction patterns observed in LEED experiments for potassium deposited on graphite \cite{li91}.

The strength of the repulsive interaction can be estimated by matching the measured pair distribution function to one generated by simulation based on a particular pair potential. Figure \ref{fig3c} shows the measured distribution function compared to the distribution for noninteracting atoms, and to distributions generated by a molecular dynamics simulation (LAMMPS) for K atoms at 11K constrained to move on a 2D plane \cite{LAMMPS}.  The simulations were perfomed assuming a dipole-dipole interaction ($V_{KK}\propto r^{-3}$), reflecting charge transfer from K that is completely screened by an image charge in the graphite \cite{div84,ter08}.  The family of simulated distributions in Fig.~\ref{fig3c} represents a range of dipole moments, always fixed in orientation to point out of the plane \cite{LAMMPS}.  As expected, a larger dipole leads to a more organized system, with sharper peaks in the pair distribution function at multiples of the nearest neighbor separation. The best fit to the experimental distribution is obtained with a dipole of 10.5$\pm 1$ D.  The data could also be fit with Coulomb interaction between monopoles ($V_{KK}\propto r^{-1}$), but the dipole interaction is more appropriate for this adatom density (typical spacing 3 nm) because the expected screening length in graphite/ene is only a few lattice spacings, less than 1 nm \cite{div84,ter08}.  The exact form of the potential is nevertheless likely to be more complicated than a simple dipole, and some deviations between simulation and experiment were observed.  For example, the pair probability below the first peak in the distribution (around 2 nm) was consistently higher in the experiment than the dipole simulation. Electron-mediated oscillatory interaction between adatoms, commonly observed on metallic surfaces \cite{rep00,kno02,zie08}, are not expected here because graphite does not support the same kind of surface states.

The value of the dipole moment for screened potassium that is extracted from these measurements can be compared to previous experiments and calculations.  Reference~\onlinecite{ost99} measured the evolution of the work function as a function of the coverage and reported a value of 9.4$\pm$1.5D, in very good agreement with our estimate. Calculations predict a value of 8.3 D at low coverage (as would be applicable here), decreasing down to 4.5 D at high coverage due to a smaller charge transfer\cite{lug07,cha08}.  EELS measurements suggest that 0.7e is transferred per adatom to graphite in the diluted phase, consistent with theoretical calculations predicting between 0.4 and 0.7e\cite{ryt07,cha08, li91b}. Using 2.7 \AA \hspace{0.5pt} as the height of the K atom above the graphite surface, 3.4 \AA \hspace{0.5pt} as the interlayer spacing, and a charge transfer of 0.7e, the dipole moment extracted from our measurements implies that the screening charge lies almost entirely in the uppermost plane\cite{ryt07}.

\begin{figure}
\subfigure{\label{fig4a}}
\subfigure{\label{fig4b}}
\subfigure{\label{fig4c}}
\subfigure{\label{fig4d}}
\includegraphics{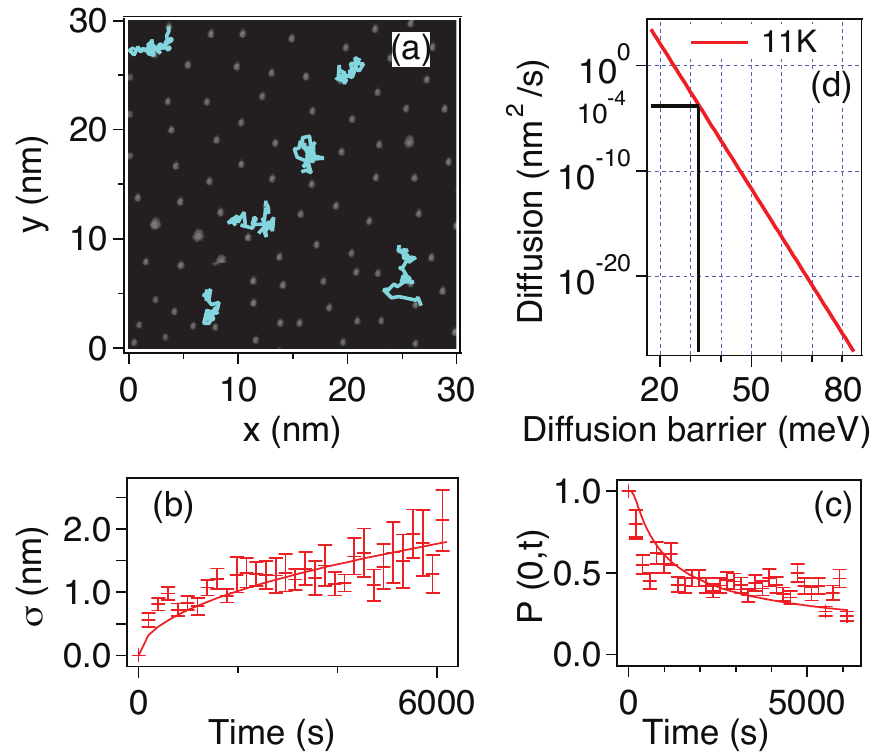}
\caption{\label{fig4}: Diffusion of potassium atoms on the graphite surface. \subref{fig4a} Initial topographic STM image and trajectory of several selected potassium adsorbates (for clarity) over 2 hours (solid lines). \subref{fig4b}  Evolution of the standard deviation $\sigma(t)$ of the distribution as a function of time. The solid line is a fit with   $\sigma(t)=\sqrt{2Dt}$, used to extract D.  \subref{fig4c} shows the maximum value of the distribution P(0,t) as a function of time with a fit with $P(0,t)=erf(\frac{r_0}{\sqrt{4Dt}})$. \subref{fig4d} Dependence of the diffusion coefficient on the diffusion barrier at 11 K.}
\end{figure}

The interaction between potassium adatoms defines their instantaneous spatial distribution statistics, but the dynamics are governed primarily by interactions with the underlying lattice.  At 11K, diffusion of K on the graphite surface was slow enough that most of the atoms remained localized while the STM tip was scanning over them, but fast enough that their their movement could be tracked over a timescale of minutes.  One notices, in  Fig.~\ref{fig2a}, that a small fraction of adatoms appear to be cut in half: these are atoms that moved to a different site as they were being imaged.  If a scan were repeated immediately after the previous frame had completed, it was observed that most of the adatoms had hopped to a new site during the course of the full scan frame. From this we can conclude that diffusion is dominated by thermal activation, and is not a tip induced artifact.

The diffusion process was characterized quantitatively by recording a time-lapse sequence of frames over two hours, with each frame acquired in a few minutes.  Figure \ref{fig4a} shows the first ($t=0$) of a sequence of scans  with K atoms at positions $\vec{r}_i(t=0)$, superimposed by trajectories, $\vec{r}_i(t)$, of several of the atoms.  Considering an ensemble of many such trajectories, one obtains a probability distribution of the displacement, $P(\delta r, t)$, with $\delta r_i(t)=|\vec{r}_i(t)-\vec{r}_i(t=0)|$, that can be compared to the expected distribution for two dimensional diffusion: 

\begin{equation}
\displaystyle P(\delta r,t)=\frac{1}{\sqrt{\pi Dt}}e^{\frac{-\delta r^2}{4Dt}},
\label{eq1}
\end{equation}

\noindent a normalized gaussian in $\delta r$ with a time-dependent standard deviation $\sigma$ for diffusion constant D. The diffusion constant was extracted from the data in two ways.  First, the measured distribution at each time was fit to a gaussian, and the resulting standard deviation was fit to  $\sigma(t)=\sqrt{2Dt}$ (Fig. \ref{fig4b}), giving $D=2.6\pm0.8 \times10^{-4}$ nm$^2$/s. Next, the experimentally measured probability of the atom to remain at its initial position was fit to $P(0,t)=erf(\frac{r_0}{\sqrt{4Dt}})$ , where $erf$ is the error function and $r_0$ the size of the bins used in the analysis (Fig.~\ref{fig4c}). This gives $D=2.3\pm0.7 \times10^{-4}$ nm$^2$/s. Combining both, we obtain $D=2.4\pm0.5 \times10^{-4}$ nm$^2$/s.

The diffusion coefficient can be related to the temperature, T, by a simple thermally activated hopping model \cite{kur97}:

\begin{equation}
\displaystyle D =\frac{k_B T n l^2}{2 h \alpha} e^{\frac{-E}{k_B T}}
\label{eq2}
\end{equation}

\noindent where k$_B$ is the Boltzman constant, $h$ is the Planck constant, $n=6$ is the number of neighbouring lattice sites available to a particle on graphite, $\alpha=2$ is the dimensionality of this system, $E$ is the energy of the diffusion barrier and $l=2.46 $ \AA\ is the spacing between lattice sites, giving $D/T= 1.5\times10^{10}  e^{-E/k_B T} $ nm$^2$s$^{-1}$K$^{-1}$ for graphite. The diffusion coefficient extracted above, D=2.4$\pm0.5\times 10^{-4}$ nm$^2$/s, at a measurement temperature T=11$\pm$1K, implies an energy barrier $E=32\pm$3 meV (Fig.~\ref{fig4d}) .  By comparison, the interaction energy between dipole moments of 10.5D at a distance of 3 nm is 2.5 meV.  One can conclude from this that lowering the temperature would lead to an organized structure of potassium atoms only for coverages significantly higher than those reported here \cite{sil04}.  This is the first experimental measurement of the hopping barrier for K on graphite, and is consistent with recent predictions of $E\approx$ 50 meV \cite{ryt07,cha08} based on density functional calculations. 

In conclusion we have reported the imaging of single K atoms adsorbed on HOPG by scanning tunnelling microscopy. We have used standard statistical methods to extract the potential landscape for a single adsorbate. It is composed of two main contributions: on the one hand an isolated potassium adsorbate will see the diffusion barrier of 32 meV on the HOPG surface; the presence of other adsorbates adds a dipole-dipole electrostatic interaction. By comparison with molecular dynamics simulations, our data suggests a value 10.5D for the screened dipole moment. This kind of local probe study could bring new insights into phase transitions occurring in two dimensions such as the ones observed for potassium adsorbed on graphite. Furthermore, the possibility to study single alkali atoms on the surface of graphite should pave the way towards the study of the local modification of the electronic structure induced by alkali atoms, which can contribute to the understanding of mechanisms driving the macroscopic phenomena observed in the graphite/alkali system.

\begin{acknowledgments}
We acknowledge J. Rottler for help with the MD simulations, and financial support by CIFAR, CFI, and NSERC. J.R. acknowledges funding from the CIFAR JF Academy.
\end{acknowledgments}


%

\end{document}